{
\documentclass{aa}
\usepackage{txfonts}
\usepackage{natbib}
\usepackage{graphicx}
\usepackage{color}
\usepackage{amssymb}
\bibpunct{(}{)}{;}{a}{}{,}

\begin{document}

\title{Determination of the far-infrared dust opacity in a prestellar core}
\author{
A. Suutarinen\inst{1,2}
\and L.K. Haikala\inst{3,2}
\and J. Harju\inst{3,2} 
\and M. Juvela\inst{2}
\and Ph. Andr{\'e}\inst{4}
\and J.M. Kirk\inst{5}
\and V. K{\"o}nyves\inst{4,6}
\and G.J. White\inst{1,7}
\fnmsep\thanks{Based partly on observations collected at the European
Southern Observatory, Chile (ESO programmes 075.C-0748 and 077.C-0562}
}

\offprints{A. Suutarinen}

\institute{
Department of Physics and Astronomy, The Open University, 
Walton Hall Milton Keynes, MK7 6AA, United Kingdom
\and 
Department of Physics, P.O.Box 64, FI-00014 University of Helsinki, Finland
\and 
Finnish Centre for Astronomy with ESO, University of Turku,
V\"{a}is\"{a}l\"{a}ntie 20, FI-21500 Piikki\"{o}, Finland
\and  
Laboratoire AIM, CEA/DSM-CNRS-Universit{\'e} Paris Diderot,
IRFU/Service d’Astrophysique, CEA Saclay, Orme des Merisiers, 91191
Gif-sur-Yvette, France
\and 
Jeremiah Horrocks Institute, University of Central Lancashire, Preston
PR1 2HE, UK
\and
Institut d'Astrophysique Spatiale, UMR8617, CNRS/Universit{\'e} Paris-Sud 11, 
91405 Orsay, France
\and
RAL Space, STFC Rutherford Appleton Laboratory, Chilton Didcot, 
Oxfordshire OX11 0QX, United Kingdom
}

\date{Received 23 February 2012 /
Accepted 11 June 2013}

\abstract {
%Text of context.  
Mass estimates of interstellar clouds from far-infrared and
submillimetre mappings depend on the assumed dust absorption
cross-section for radiation at those wavelengths.  }
{%Text of aims 
The aim is to determine the far-IR dust absorption cross-section in a
starless, dense core located in Corona Australis.  The value is needed
the determining of the core mass and other physical
properties. It can also have a bearing on the evolutionary stage of the
core. }
{%Text of methods
  We correlated near-infrared stellar $H-K_{\rm s}$ colour excesses of
  background stars from NTT/SOFI with the far-IR optical depth
  map, $\tau_{\rm FIR}$, derived from \textit{Herschel} 160, 250, 350,
  and 500 $\mu$m data.  The \textit{Herschel} maps were also used to
  construct a model for the cloud to examine the effect of temperature
  gradients on the estimated optical depths and dust absorption
  cross-sections.  }
 {%Text of results
   A linear correlation is seen between the colour $H-K_{\rm s}$ and 
   $\tau_{\rm FIR}$ up to high extinctions ($A_V \sim 25$). The correlation 
   translates to the average extinction ratio 
   $A_{250\mu{\rm m}}/A_J = 0.0014 \pm 0.0002$, assuming a standard 
   near-infrared extinction law and a dust emissivity index $\beta=2$. 
   Using an empirical $N_{\rm H}/A_J$
   ratio we obtain an average absorption cross-section per H nucleus
   of $\sigma^{\rm H}_{250\mu\rm m} = (1.8 \pm 0.3) \times 10^{-25}$ cm$^{2}$
   H-atom$^{-1}$, corresponding to a cross-section per unit mass of
   gas $\kappa_{250\mu{\rm m}}^{\rm g} = 0.08 \pm 0.01 $ cm$^{2}$g$^{-1}$.
   The cloud model however suggests that owing to the bias caused by
   temperature changes along the line-of-sight these values
   underestimate the true cross-sections by up to 40\% near the centre
   of the core.  Assuming that the model describes 
   the effect of the temperature variation on $\tau_{\rm FIR}$ correctly, we find
   that the relationship between $H-K_{\rm s}$ and $\tau_{\rm FIR}$ agrees with 
   the recently determined relationship between $\sigma^{\rm H}$ and 
   $N_{\rm H}$ in Orion A.
  }
  {%Text of conclusions
    The derived far-IR cross-section agrees with previous
    determinations in molecular clouds with moderate column densities,
    and is not particularly large compared with some other cold cores.
    We suggest that this is connected to the core not beng very dense 
    (the central density is likely to be $\sim 10^5$
    cm$^{-3}$) and judging from previous molecular line data, it
    appears to be at an early stage of chemical evolution. }

\maketitle

\section{Introduction}

Determining the emission properties of interstellar dust is useful not
only for providing reliable molecular cloud mass estimates from thermal dust
emission maps, but also for testing ideas about dust evolution. In
several previous studies the far-infrared to visual or far-IR to
near-IR extinction ratio ($A_\lambda/A_V$ or $A_\lambda/A_J$),
sometimes called the {\sl emissivity}, has been studied by combining
near-infrared photometry and dust continuum maps
(\citealt{2003A&A...399L..43B}; \citealt{2003A&A...399.1073K};
\citealt{2005ApJ...632..982S}; \citealt{2007A&A...466..969L};
\citealt{2011ApJ...728..143S}). One of the results from these studies is
that the dust emissivity changes from region to region, and it shows a
tendency to increase with decreasing dust temperature 
\citep{2012A&A...541A..19A}. 
This trend agrees well with theoretical predictions, is thought
to be caused primarily by dust coagulation in the cold, dense cores of
molecular clouds, and is further accentuated by ice formation on the
surfaces of dust grains (\citealt{1994A&A...291..943O};
\citealt{2003A&A...398..551S}; \citealt{2009A&A...506..745P};
\citealt{2011A&A...532A..43O}).

The availability of multi-frequency data from the \textit{Planck} and
\textit{Herschel} satellites has vastly improved the accuracy of the
determination of dust properties and physical conditions in molecular
clouds. In particular, the wavelengths accessible to 
\textit{Herschel} cover both sides of the emission maximum of
cold ($T\sim 10$ K) interstellar dust.  Earlier surveys usually had
only two or three frequencies in the submillimetre/far-infrared and
often only on one side of the peak.

In this paper we determine the far-infrared dust emissivity and
absorption cross-section in a prestellar core using sensitive
near-infrared photometry in the $H$ and $K_{\rm s}$ bands in conjunction with
\textit{Herschel} maps at four wavelengths. In this way we believe we
can achieve higher accuracy than previous studies have. 
Furthermore, the target core provides a useful
reference because it has been claimed that it represents a very early stage
of chemical evolution where molecular depletion is not significant
\citep{2003cdsf.conf..331K}. The core, which lies in the ``tail'' of the
Corona Australis molecular cloud, can be found in the Planck Early
Release Compact Source Catalogue as PLCKECC G359.78-18.34\footnote{\tt
  http://irsa.ipac.caltech.edu/applications/planck/}, and has been
previously called '\object{CrA C}' \citep{1993A&A...278..569H} and '\object{SMM 25}'
\citep{2003A&A...409..235C}. So far, no indication of star formation
taking place in CrA C has been found in the millimetre continuum
\citep{2003A&A...409..235C}, in submillimetre (Herscel), or in the
mid- and near-infrared \citep{2011ApJS..194...43P}.

In Sect.~2 we briefly discuss some aspects of $JHK$ photometry towards
highly reddened stars, and in Sect. 3 we describe the near- and
far-infrared observations used in the present work. In Sect.~4, the
near-IR reddening is correlated with the far-IR optical depth, and
the result is used to derive the dust absorption
cross-section. Finally, in Sect. 5 we discuss the significance of the
obtained results.
 
\section{On $JHK$ photometry at high extinctions}

Analysis of $JHK$ near-infrared photometry involves some
uncertainties and complications.  The $JHK$ system is not uniquely
defined because different detectors and filter sets have been
used. Additionally the transmissions in the $JHK$ bands depend
strongly on the atmospheric transmission (atmospheric H$_2$O absorption
lines) and thus vary depending on the observing site and is even
subject to night-to-night variations.  As a consequence one should be
careful when comparing photometric data obtained at different epochs
and observing sites. This is also true when adapting a $JHK$ reddening
slope because these slopes depend on the particular filter-detector
combination.  The modified $K_{\rm s}$ is narrower than the $K$ filter
and excludes some of the strong atmospheric lines and is thus less
influenced by the atmospheric opacity.

The sensitivity of near-infrared observations has increased
dramatically during the last decade, and it is now possible to
observe routinely highly reddened, faint stars. The reddening
affects the observed object spectral energy distribution (SED) and for
highly reddened sources the effective wavelength of the filter shifts
towards longer wavelengths. This introduces a new effect as the $JHK$
reddening slope changes depending on the reddening. This phenomenon
was already noted in the $UBV$ system \citep[see
e.g.][]{1974ASSL...41.....G}, and has been later studied e.g. using
the Two Micron All Sky Survey (2MASS)\footnote{\tt
  http://irsa.ipac.caltech.edu/Missions/2mass.html}
\citep{2008BaltA..17..277S}, and the UKIRT Infrared Deep Sky Survey
(UKIDSS) data \citep{2009MNRAS.400..731S}. As noted in
\cite{2009MNRAS.400..731S} one should refer to a reddening {\sl track}
instead of a reddening slope. The importance of the photometric system
is highlighted by the fact that in the 2MASS system the reddening
track of highly reddened giants curves down in the $JHK$ colour-colour
diagram whereas in the MKO system (UKIDSS) the track curves up. In
addition to reddening this effect depends on the spectral type of the
source. The effect is noticeable for $H-K$ indices larger than $\sim
1.5$.  In the present work the highest $H-K_{\rm s}$ indices are nearly 3.
 
\section{Data acquisition}
\subsection{NTT/SOFI  observations}
\label{sect:nirobservations}

The dense core region of CrA~C was imaged in $J$, $H$, and $K_{\rm s}$
  using the Son Of Isaac (SOFI) near-infrared instrument on the New
  Technology Telescope (NTT) at the La Silla observatory in May 2005
  and July 2006. The SOFI field of view is 4\farcm9 and the pixel
size is 0\farcs288. The observing was done using the standard
jittering mode in observing blocks of approximately 1 hour.  
Standard stars from the faint NIR standard list of
  \citet{1998AJ....116.2475P} were observed frequently during the
  nights.
Because very few stars were seen in the first $J$ band
observing block the observations were continued only in the $H$ and
$K_{\rm s}$ bands.  The total on source time was 40 min, 104 min and 540 min
in the $J$, $H$, and $K_{\rm s}$ filters, respectively.  The average seeing
was $\sim$0\farcs7.

The IRAF{\footnote {IRAF is distributed by the National Optical
    Astronomy Observatories, which are operated by the Association of
    Universities for Research in Astronomy, Inc., under cooperative
    agreement with the National Science Foundations}} external package
XDIMSUM was used to reduce the imaging data.  The images were searched
for cosmic rays and sky-subtracted. The two nearest 
  background images in time to each image were used in the sky
subtraction. An object mask was constructed for each image. Applying
these masks in the sky subtraction produced hole masks for each
sky-subtracted image.  Special dome flats and illumination correction
frames provided by the NTT team were used to flat field and to
illumination correct the sky-subtracted SOFI images.  Rejection masks
combined from a bad pixel mask and individual cosmic ray and hole
masks were used when averaging the registered images.  The coordinates
of the registered images were derived from the 2MASS catalogue
\citep{2006AJ....131.1163S}.

The SExtractor software v 2.5.0 \citep{1996A&AS..117..393B} package was
used to obtain stellar photometry of the reduced SOFI images. 
Galaxies were excluded using the SExtractor CLASS keyword and by
visual inspection of the images.  

  The zero points of the coadded data  were fixed
  using standard star measurements and checked using stars in
  common with the less deep HAWK-I imaging of CrA reported in 
  \citet{2011ApJ...736..137S} (see below).  The limiting magnitudes (for a
  formal error of 0.1 magnitudes) are approximately 21\fm5, 20\fm5, and
  20\fm8 for $J$, $H$, \ and $K_{\rm s}$, respectively.  After elimination of
  the nonstellar sources 25, 266, and 369 stars remained in $J$, $H$,
  and $K_{\rm s}$, respectively. In the subsequent analysis we use the
  266 stars detected both in the $H$ and $K_{\rm s}$ filters. The locations
  of these stars are indicated with blue plus signs in
  Fig.~\ref{figure:taumap}. As one can see in this Figure, no stars
  were detected in the very centre of CrA~C. All detections lie
  at the edges of the core.

  The extracted $JHK_{\rm s}$ magnitudes were converted into Persson
  magnitudes, and from these to the 2MASS photometric system as
  described in \cite{2007A&A...466..137A}.  However, these conversion
  formulae could not be directly applied to all of the SOFI data
  because for most stars $J$ band magnitudes were not available owing
  to the high extinction. Using the equations (1)-(6) of
  \cite{2007A&A...466..137A} the conversion formula for the $(H-K_{\rm s})$
  index from the instrumental magnitudes to the 2MASS system can be
  written as
  \begin{equation}
  (H-K_{\rm s})_{\rm 2MASS} = 1.019 (H-K_{\rm s}) - 
  0.046(J-K_{\rm s}) + 0.005 \; ,
  \label{eq:HminusK_2MASS} 
  \end{equation}
  where the magnitudes on the right are SOFI magnitudes. 
  We have attempted to decrease the systematic error arising from the missing 
  $J$ magnitude by estimating the $(J-K_{\rm s})$ colour using the observed 
  $H-K_{\rm s}$ index, the intrinsic $J-K$ and $H-K$ colours of 
  giant stars from \cite{1988PASP..100.1134B}, and the extinction ratios 
  $A(\lambda)/A(K_{\rm s})$ derived by \cite{2005ApJ...619..931I}. These give 
  \begin{equation}
  \begin{array}{lcl}
  (J-K_{\rm s})_{\rm est} &=& (J-K_{\rm s})_0 - 
                         \frac{E(J-K_{\rm s})}{E(H-K_{\rm s})}(H-K_{\rm s})_0  +
                         \frac{E(J-K_{\rm s})}{E(H-K_{\rm s})} (H-K_{\rm s}) 
                          \\
                     & & \\

                       &=& (0.39\pm0.11) + (2.72\pm0.48)\times(H-K_{\rm s}) \; .
  \end{array}
  \label{eq:JminusK_est}  
  \end{equation}

  The ratio of colour excesses $E(J-K_{\rm s})/E(H-K_{\rm s}) =
  2.72\pm0.48$ is obtained using the average $A(J)/A(K_{\rm s})$ and
  $A(H)/A(K_{\rm s})$ values listed in Table 1 of
  \cite{2005ApJ...619..931I}. The corresponding values of 
  $(J-K)_0
  - \frac{E(J-K_{\rm s})} {E(H-K_{\rm s})} (H-K)_0$ range from $0.27$
  to $0.50$ for giants of the types G0 to M7 using the
  intrinsic colours given in \cite{1988PASP..100.1134B}. The possible
  variation of the intrinsic colours combined with the uncertainty of
  the colour excess ratio causes an additional uncertainty of $\sim
  0\fm03$ to the derived $H-K_{\rm s}$ index. 
This small uncertainty is added in quadrature to the uncertainties
  of $H-K_{\rm s}$ owing to formal photometric errors ($\sim
  0\fm09$) and the uncertainty in the magnitude
  zero-point ($\sim 0\fm07$ for $H-K_{\rm s}$).
 
The conversion to the 2MASS system using Eqs.~\ref{eq:HminusK_2MASS}
and \ref{eq:JminusK_est} reduces the instrumental $H-K_{\rm s}$
indices by about 10 percent.  The observed $H-K_{\rm s}$ indices for the
  highly reddened stars in our sample range from 1\fm5 to 2\fm9, and
  for these stars the correction lies between $-0\fm14$ and
  $-0\fm25$. Ignoring the colour-dependent term in the conversion to
the 2MASS system would bias systematically the $H-K_{\rm s}$ indices
used in the following analysis.

\subsection{VLT/HAWK-I observations}

The core CrA C was partially covered by two $7\farcm5 \times 7\farcm5$
fields taken with the HAWK-I IR camera on the VLT as part of the
survey of Corona Australis reported in \citet{2011ApJ...736..137S}
(ESO program 083.C-0079). A more detailed description of the
observations and reduction procedures can be found in that paper.
The photometry zero point of the HAWK-I data was fixed to 2MASS
objects present in the fields. The typical calibration errors range
from 1\% to 5\%, depending on the weather conditions. Therefore,
signal to noise and background remain the main sources of uncertainty,
especially for the fainter objects. The data are complete in the
following dynamical ranges: $J = 11-18.5$ mag, $H = 11-18.5$ mag, $K =
10.5-18$ mag. The locations of HAWK-I stars with reasonably small
errors ($<0\fm1$ in $H$ and $K$) lying within a radial distance of
5\farcm5 from the core centre are marked with red crosses in
Fig.~\ref{figure:taumap}. The total number of these stars is 964.

\begin{figure}[htb]
  \resizebox{\hsize}{!}{\includegraphics{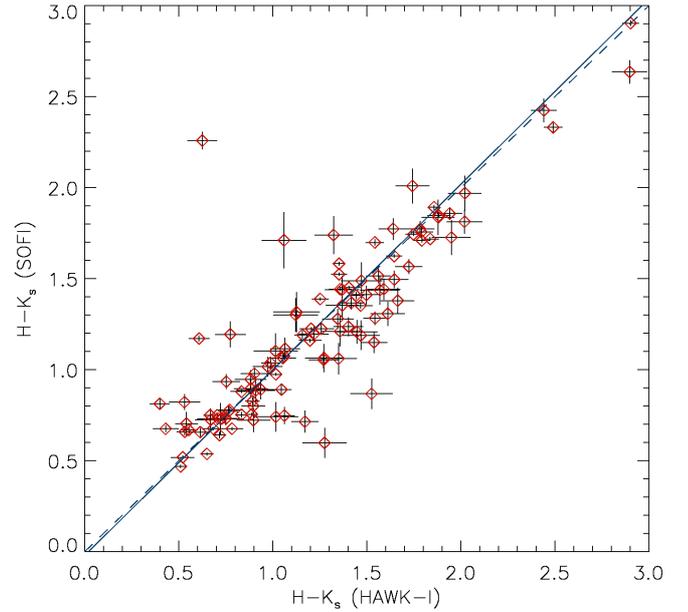}}
  \caption{Relationship between the $H-K_{\rm s}$ colours of the 109 stars
    detected with both SOFI and HAWK-I, before performing the
    colour-dependent conversion to the 2MASS system. The photometric
    errors of the data points are indicated. The solid blue line
    shows the fit to the data, and the dashed line represents the one-to-one 
    correlation.}
\label{figure:H-K_SOFI_vs_HAWK-I} 
\end{figure}

  Of the 266 stars in our SOFI sample 109 were detected with
  HAWK-I. In Fig.~\ref{figure:H-K_SOFI_vs_HAWK-I} we correlate the
  $H-K_{\rm s}$ indices of stars common to both surveys. Only a zero-point
  correction to the 2MASS system has been done to the magnitudes
  presented in this figure. This is because no colour-dependent
  correction has been applied  to the HAWK-I data (Aurora
  Sicilia-Aquilar, private communication) as the characteristics of
  the HAWK-I filters needed for this correction are not available. One
  can see the uncorrected SOFI colours agree with the HAWK-I data.
  After the colour correction for SOFI, the slope of the
  correlation deviates clearly from unity because this correction
  reduces the $H-K_{\rm s}$ indices by about 10\%. Because our main purpose
  is to derive the dust opacity towards the core where the colour 
  correction has a noticeable effect, we have decided not to use the
  HAWK-I data in the subsequent analysis.

\subsection{2MASS data}

Of the stars detected in the 2MASS survey 58 (again, including only
those with $\Delta K$ and $\Delta H$ $<0\fm1$) lie within a distance
of $5\farcm5$ from the core centre . The positions of these stars
(excluding those coincident with HAWK-I or SOFI stars) are marked in
Fig.~\ref{figure:taumap}.  Only two of the 2MASS
stars coincide with those in the SOFI sample.  The limiting magnitudes
of the 2MASS data are $15\fm8$, $15\fm1$, and $14\fm2$ in $J$, $H$,
and $K$, respectively. Because the 2MASS data clearly probe the
diffuse envelope around the core we will not use them in the
derivation of the submillimetre opacity.

\subsection{Herschel data}

\begin{figure}
\resizebox{\hsize}{!}{\includegraphics{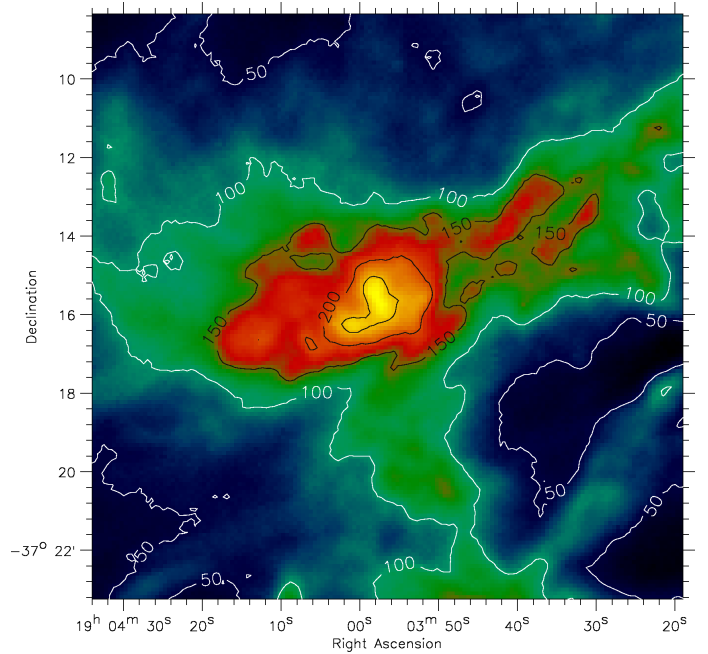}}
\caption{Herschel $\lambda=250\mu$m map of CrA C. 
The contours indicate the intensity $I_{250\mu{\rm m}}$ levels ranging from 
50 to 250 MJy/sr. The FWHM of the Herschel beam at this wavelength is
18\arcsec.}
\label{figure:I250map}
\end{figure}

\begin{figure}
\resizebox{\hsize}{!}{\includegraphics{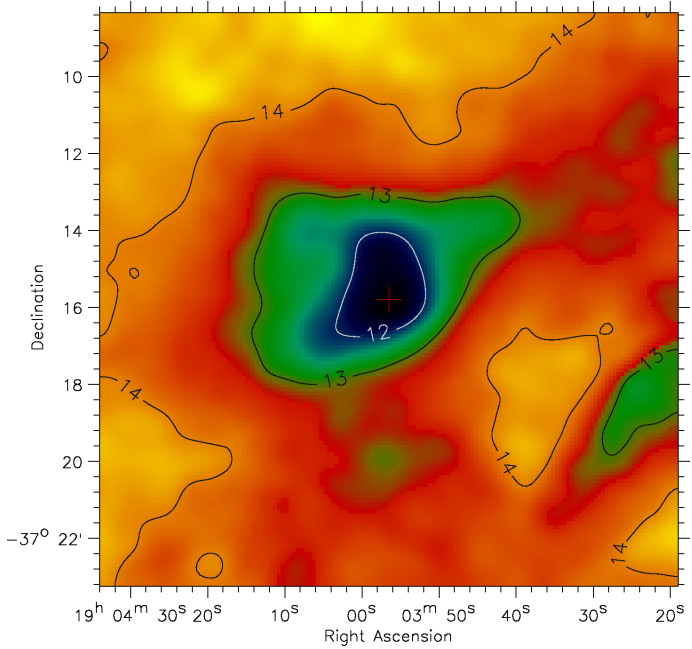}}
\caption{Dust temperature ($T_{\rm D}$ in K) map of CrA C derived from 
Herschel data.}
\label{figure:tdustmap}
\end{figure}

The $15\arcmin\times15\arcmin$ far-IR and submillimetre maps of CrA C
used in this study were extracted from an extensive mapping of the
Corona Australis region made as part of the \textit{Herschel} Gould Belt
Survey \citep{2010A&A...518L.102A} with the SPIRE
\citep{2010A&A...518L...3G} and PACS \citep{2010A&A...518L...2P}
instruments onboard \textit{Herschel} \citep{2010A&A...518L...1P}.  The
original CrA maps cover a region of about $5\degr\times2\degr$ at the
wavelengths 70 and 160 $\mu$m (PACS) and 250, 350, and 500 $\mu$m (SPIRE).
The SPIRE observations were reduced with HIPE 7.0, using
modified pipeline scripts, then, the default naive mapping routine was
applied to produce the final maps.  PACS data was processed within
HIPE 8.0 up to level 1 after which we employed Scanamorphos v16
\citep{2012arXiv1205.2576R} to create the map
products. The maps of the whole CrA field will become available on the
\textit{Herschel} Gould Belt Survey Archives\footnote{\tt
  http://gouldbelt-herschel.cea.fr/archives}. \rm The beam FWHMs for
each wavelength are 36$\arcsec$, 25$\arcsec$, 18$\arcsec$,
12$\arcsec\times$16$\arcsec$, and 6$\arcsec\times$12$\arcsec$ for 500,
350, 250, 160, and 70 $\mu$m, respectively, the PACS beams being
clearly elongated in observations carried out in the parallel
mode. The \textit{Herschel} intensity map at $\lambda=250\mu$m, $I_{250\mu{\rm
    m}}$, is shown in Fig.\ref{figure:I250map}.

Based on comparison with the Planck satellite maps of the region the
following offsets were added to the \textit{Herschel} maps from 70 to 500
$\mu$m, respectively: 2.5, 14.9, 15.6, 8.7, 4.0 MJy/sr. The
maps at 160, 250, 350, and 500 $\mu$m were convolved to a resolution
of $40\arcsec$ (FWHM), and the intensity distributions were fitted
with a modified blackbody function, $I_\nu \approx B_\nu(T_{\rm
  dust})\tau_\nu \propto B_\nu(T_{\rm dust}) \, \nu^{\beta}$, which
characterises optically thin thermal dust emission at far-IR and
submillimetre wavelengths.  We fixed the emissivity/opacity exponent
to $\beta=2.0$. The resulting $T_{\rm dust}$ map is shown in
Fig.~\ref{figure:tdustmap}. The distribution of the optical depth,
$\tau(250\mu{\rm m}) = I_{250\mu{\rm m}}/B_{250\mu{\rm m}}(T_{\rm
  dust})$, is shown in Fig.~\ref{figure:taumap}. On this map we have
plotted the SOFI stars selected for extinction
estimates (see below), and the 2MASS and HAWK-I stars within a radius of 
$5\farcm5$ of the core centre.
 
The errors of $\tau_{\rm 250\mu m}$ were estimated from a Monte Carlo
method using the $1\sigma$ error maps provided for the three SPIRE
bands and a 7\% uncertainty of the absolute calibration for all four
bands according to the information given in SPIRE and PACS manuals 
(SPIRE Observers'  Manual, Version 2.4; PACS Observer's Manual, Version 2.4)
Different realisations of the $\tau_{\rm 250\mu m}$ map were
calculated by combining the four intensity maps with the corresponding
error maps, assuming that the error in each pixel is normally
distributed. The $\tau_{\rm 250\mu m}$ error in each pixel was
obtained from the standard deviation of one thousand realisations.

\begin{figure*}[htb]
\resizebox{\hsize}{!}{
\includegraphics{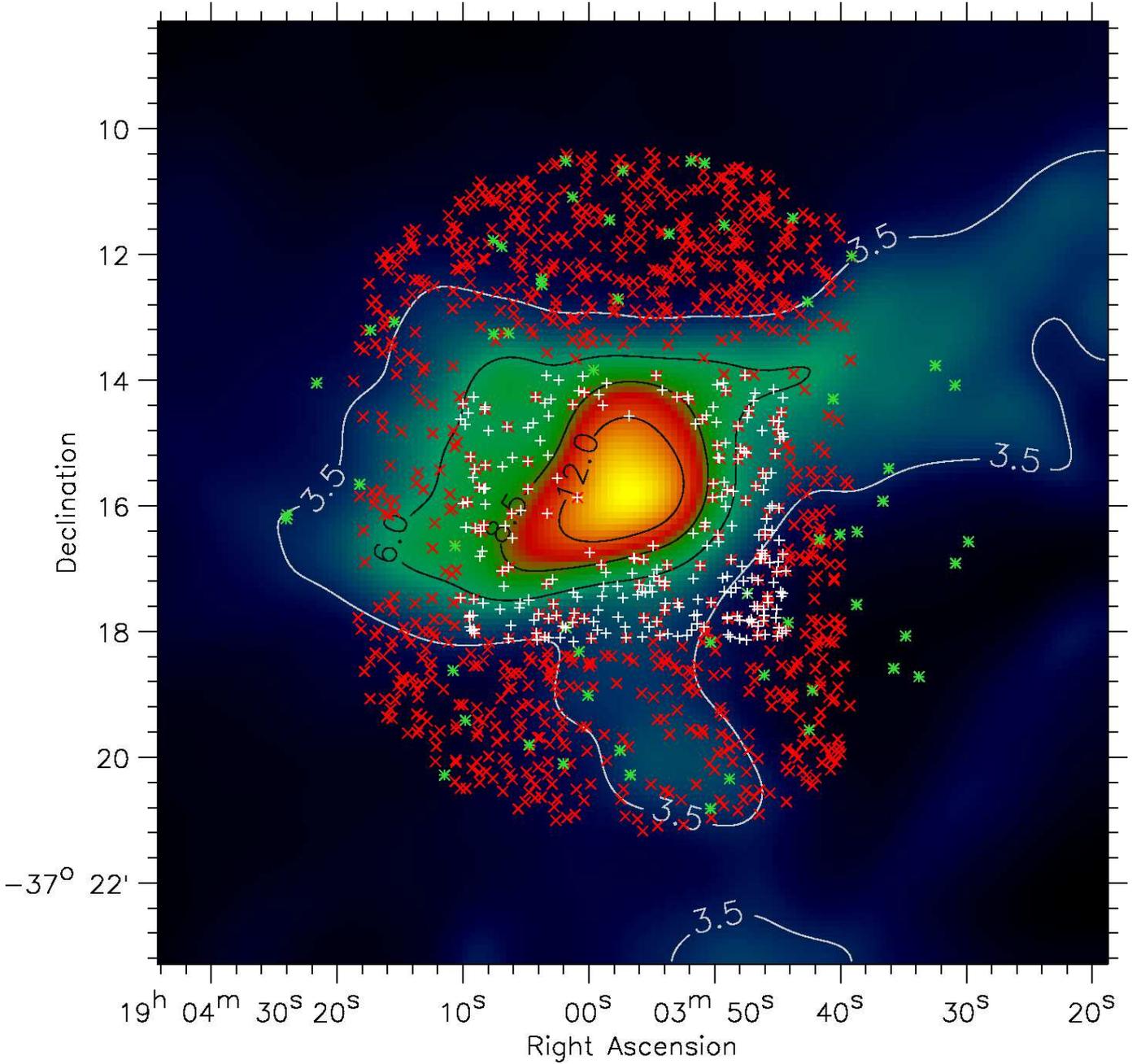}
}
\caption{Locations of background stars on the optical depth
  $\tau(250\mu{\rm m})$ map from \textit{Herschel}. The contour
    labels give $\tau(250\mu{\rm m})$ values multiplied by 1000.
  Different symbols have been used to distinguish between stars from
  the three near-IR surveys: 2MASS (green asterisks), HAWK-I (red
  crosses), and SOFI (white plus signs). Of the HAWK-I and 2MASS
    stars only those within a $5\farcm5$ radius of the map centre are
    shown here.}
\label{figure:taumap}
\end{figure*}

\section{Data analysis and results}

As we are dealing with a dense core with high near-infrared extinction
the number of background stars with accurate $J$ magnitudes is low. To
obtain reasonably good statistics, and to extend extinction estimates
close to the core centre, we have only used the $H$ and $K_{\rm s}$
bands where the extinction is less severe than in $J$. A further
motivation for the use of $H$ and $K_{\rm s}$ is the fact that the intrinsic
$H-K$ colours of giants, the class of the background stars most
probably observed towards the cloud span a rather
narrow range of $(H-K)_0 = 0.07-0.31$ mag \citep{1988PASP..100.1134B}.
The stellar near-infrared colour excess $E(H-K)\equiv (H-K)-(H-K)_0$
can be used to evaluate the dust column density \citep{ladalada1994}.

Because of the reasons discussed in Sects.~2.1-2.3, the HAWK-I and
2MASS stars were used only for checking the zero point of the magnitude
scale for the SOFI stars, but not in the estimation of the dust
opacity in the CrA~C core. For this purpose we used SOFI stars with
photometric $1-\sigma$ errors of $\Delta H < 0.1$ mag, $\Delta K <
0.1$ mag. The number of stars fulfilling these criteria is 266. The
locations of these stars are indicated with white plus signs in the
\textit{Herschel} far-IR optical depth ($\tau_{250\mu{\rm m}}$)
map shown in Fig.~\ref{figure:taumap}.

\begin{figure}[t]
  \resizebox{\hsize}{!}{\includegraphics{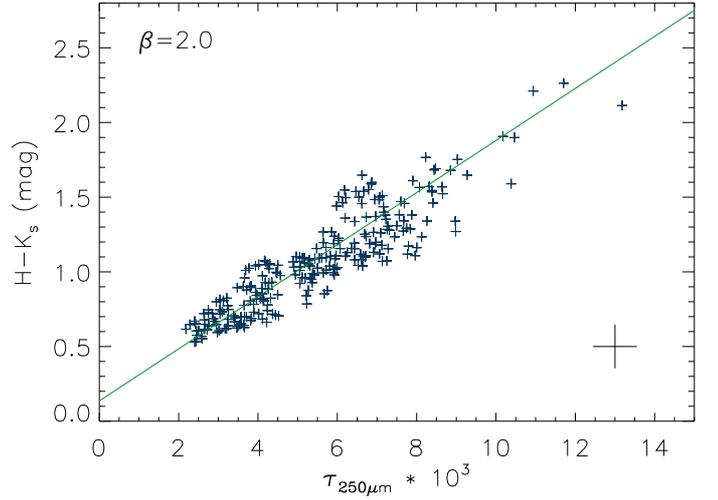}}
  \caption{$H-K_{\rm s}$ colours of the background stars detected with SOFI 
    as a function of
    far-IR optical depth of the dust emission. 
   The mean error of the data points is indicated with a cross 
   in the bottom right.
     }
\label{figure:H-Kvstau} 
\end{figure}
 
Besides the colour excess $E(H-K_{\rm s})$ also the optical depth
$\tau_{250\mu{\rm m}}$ measures the column density of dust, albeit at
very different angular resolution. We have tried to reconcile this
difference in the following way: From the \textit{Herschel} map we
have only taken those pixels that contain one or more background
stars. The $H-K_{\rm s}$ value assigned to this pixel is then
calculated as the average of the neighbouring stars weighted by a
$40\arcsec$ Gaussian beam to match the resolution of the smoothed dust
emission maps. The $H-K_{\rm s}$ measurements for individual stars
taken in the averages were furthermore weighted according to their
errors owing to the photometric noise and to the uncertainty of the
$J-K_{\rm s}$ colour needed for the conversion to the 2MASS system as
discussed in Sect.~\ref{sect:nirobservations}. The errors of the
$H-K_{\rm s}$ values corresponding to pixels in the $\tau_{250\mu{\rm
m}}$ map were calculated using the standard formula for a weighted
average.  The resulting correlation plot $H-K_{\rm s}$
vs. $\tau(250\mu{\rm m})$ is shown in Fig.~\ref{figure:H-Kvstau}.  The
typical (mean) error for the colours of the SOFI stars is
indicated. The horizontal error bar represents the mean error of the
$\tau_{250\mu{\rm m}}$ estimates (see Sect. 2.4).

A linear fit to the  SOFI data  gives the following model (shown in Fig.~\ref{figure:H-Kvstau}):
$$
H-K_{\rm s} =  (0.14 \pm 0.02)  + (174 \pm 4)  \, \tau_{250\mu{\rm m}} \; .
$$ 
The point where the model intersects the y-axis should represent
  the average intrinsic colour $(H-K_{\rm s})_0$ of the background
  stars.  This assumption seems reasonable since the intersection,
  0\fm14, corresponds to the intrinsic $H-K$ colour of a K type giant
  \citep{1988PASP..100.1134B}. Therefore, converting the far-infrared
  optical depth to extinction, the fit result can be written as
  $A(250\mu{\rm m}) = (0.0062\pm0.0001)\,E(H-K_{\rm s})$. Here we have
  used the relationship $A_{250\mu{\rm m}}= 1.086\,{\tau _{250\mu{\rm
        m}}}$.

The colour excess can be converted to the near-infrared extinction, 
e.g. $A(J)$, using the wavelength dependence of extinction. 
According to the extinction ratios, 
$A(J)/A(K_{\rm s})=2.50\pm0.15$ and 
$A(H)/A(K_{\rm s})=1.55\pm0.08$, 
presented in Table 1 of \cite{2005ApJ...619..931I} the relationship between
$A(J)$ and $E(H-K_{\rm s})$ is $A_J = (4.55\pm0.72) E(H-K_{\rm s})$.
This relationship is consistent with the extinction law from
\cite{1989ApJ...345..245C}, when the wavelengths of the 2MASS $J$,
$H$, and $K_{\rm s}$ bands are used in their interpolation formula
(2a) and (2b), giving $A_J = 4.44 E(H-K_{\rm s})$. 
Using the results of  \cite{2005ApJ...619..931I} we obtain the following
extinction ratio:
$$
\frac{A_{250\mu{\rm m}}}{A_J} = 0.0014 \pm 0.0002 \; . 
$$
This extinction ratio is very similar to the value 0.0015
listed in Table~1 of \cite{1990ARA&A..28...37M}, and also agrees  
with the synthetic extinction law based on the dust model developed 
by \cite{2001ApJ...548..296W} and 
\cite{2001ApJ...554..778L}\footnote{data available at \\
{\tt http://www.astro.princeton.edu/\~draine/dust/dustmix.html}}.
According to the assumption $\beta=2.0$ used here, the extinction
ratio for other wavelengths longer than 160 $\mu$m can be obtained by
multiplying the number above by $(250\mu{\rm m}/\lambda)^2$.
 
To facilitate comparison with some previous results we note that,
  extrapolating the extinction ratio to the submillimetre regime, our
  result corresponds to $A_{850\mu{\rm m}}/A(K_{\rm s})=3.0\times
  10^{-4}$ or $A_{850\mu{\rm m}}/A_V=3.9\times 10^{-5}$. In the latter
  ratio we have assumed that $A_J/A_V=0.333$
  (\citealt{1989ApJ...345..245C} for $R_V=5.5$).

\vspace{2mm}

The dust absorption cross-section per H nucleus, $\sigma^{\rm
  H}_{250\mu{\rm m}}$, or per unit mass of gas, $\kappa_{250\mu{\rm
    m}}^{\rm g}$, can be estimated using empirical determinations of
the $N_{\rm H}/A_J$ or $N_{\rm H}/A_V$ ratio.  As discussed by
\cite{2012ApJ...751...28M}, the ratios of near-infrared colour excess,
or alternatively near-infrared extinction to $N_{\rm H}$ are likely to
change less than for example the ratio of $A_V$ to $N_{\rm H}$ when
dust grains evolve in dense material. Therefore one could expect that 
the ratio $N_{\rm H}/A_J$ does not change substantially from diffuse to 
dense clouds. 

The $N_{\rm H}/E(B-V)$ ratio from the Ly$\alpha$ absorption
measurement of \citet{1978ApJ...224..132B} with the value $R_V=3.1$
characteristic of Galactic diffuse clouds and the corresponding
$A_J/A_V$ ratio ($=0.282$) give $N_{\rm H}/A_J = 6.6\times10^{21} \,{\rm
  cm}^{-2} {\rm mag}^{-1}$.  \citet{2003A&A...408..581V} determined in
the dark cloud $\rho$ Oph $N_{\rm H}/A_J = 5.6-7.2 \times 10^{21} \,{\rm
  cm}^{-2} {\rm mag}^{-1}$ (depending on the assumed metal abundances)
by combining X-ray absorption measurements with near-infrared
photometry.  Recently, \citet[][Eq. 9]{2012ApJ...751...28M} compiled a
correlation between $N_{\rm H}$ and $E(J-K_{\rm s})$ from several
previous surveys. Their best-fit slope, 
$N_{\rm H}/E(J-K_{\rm s})= (11.5\pm0.5)\times 10^{21}\; 
{\rm cm^{-2}}\,{\rm mag}^{-1}$, 
together with $A_J/E(J-K_{\rm s}) = 1.67\pm0.07$ from 
\cite{2005ApJ...619..931I} imply the ratio  
$N_{\rm H}/A_J=(6.9\pm0.4)\times10^{21} \,{\rm cm}^{-2} {\rm mag}^{-1}$. 
Adopting this $N_{\rm H}/A_J$ ratio we
estimate for $\sigma^{\rm H}$ and $\kappa^{\rm g}$ the following values:
$$
\begin{array}{lll}
\sigma^{\rm H}_{250\mu{\rm m}} &=&  (1.8\pm 0.3)\times 10^{-25} \, 
                                    {\rm cm}^2 \, {\rm (H \, atom)}^{-1} 
\; , \\
  & & \\

\kappa_{250\mu{\rm m}}^{\rm g} &=& (0.08\pm0.01) \; {\rm cm}^2 {\rm g}^{-1}
                \, \mbox{of gas} \; ,
\end{array}
$$
where $\kappa^{\rm g}$ is obtained from $\sigma^{\rm H}$ by dividing
it by the average gas particle mass per H nucleus ($ = 1.4 \, m_{\rm H}$, 
assuming 10\% He). The range of $N_{\rm H}/A_J$ values obtained
by \citet{2003A&A...408..581V} would imply
$ \sigma^{\rm H}_{250\mu{\rm m}} = 1.8-2.3\times 10^{-25} \, {\rm cm}^2 \, {\rm H}^{-1}$, 
$\kappa_{250\mu{\rm m}}^{\rm g} = 0.08-0.10 \; {\rm cm}^2 {\rm g}^{-1}$.
Note that $\kappa$ is often given per unit mass of dust in which case
the number above should be multiplied by the assumed gas-to-dust mass
ratio (usually 100-150).

The values implied by the parameter
$C_{250}$ given in \citet{1983QJRAS..24..267H} are $\sigma^{\rm
  H}_{250\mu{\rm m}} = 1.4 m_{\rm H}/C_{250} = 2.3\times 10^{-25} \, {\rm
  cm}^2 {\rm (H\, atom)}^{-1}$, $\kappa_{250\mu{\rm m}}^{\rm g}
=1/C_{250} = 0.10\,{\rm cm^{2}\, g^{-1}}$, that is, slightly larger 
than those obtained above. Again, $\kappa_{250\mu{\rm m}}^{\rm g}$ can be
converted to other wavelengths by multiplying by $(250\mu{\rm m}/\lambda)^2$. 
For $160\mu{\rm m}$ we obtain  $\kappa_{160\mu{\rm m}}^{\rm g} = (0.19\pm0.03)
 \, {\rm cm}^2 {\rm g}^{-1}$, and extrapolating our value to $850\mu{\rm m}$ we get $\kappa_{850\mu{\rm m}}^{\rm g} = (0.007\pm 0.001) \, {\rm
  cm}^2 {\rm g}^{-1}$.

\begin{figure}
\resizebox{\hsize}{!}{
\includegraphics{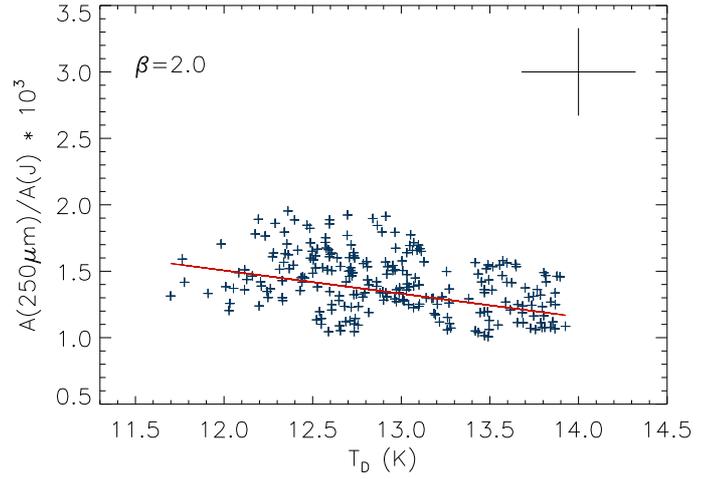}
}
\caption{Extinction ratio $A_{250\mu{\rm m}}/A_J$ vs. $T_{\rm
    dust}$. The red line represents a linear fit to the data points
  using SOFI - \textit{Herschel} comparison (blue crosses). The mean error 
of the data points is indicated with a cross in the top right.}
\label{figure:emiss_vs_td}
\end{figure}

The ratio $A_{250\mu{\rm m}}/A_J$ is plotted as a function of $T_{\rm
  dust}$ in Fig.~\ref{figure:emiss_vs_td}. One can see that the ratios
  show a large scatter, but a decreasing tendency towards higher
  temperatures can be discerned. A linear fit to the data points
  gives the following relationship $A_{250\mu{\rm
  m}}/A_J=(3.6\pm0.5)\times 10^{-3}-(1.7\pm0.4)\times 10^{-4}\,T_{\rm
  dust}$, which suggests that the average ratio increases from $0.0012$
  to $0.0016$ (30\%) when $T_{\rm dust}$ decreases from 14 K to 12 K.
  Using the previously quoted $N_H/A_J$ ratio $6.9\times 10^{21} \,{\rm
  cm}^{-2} {\rm mag}^{-1}$ this tendency in the extinction ratio can
  be translated to the dust extinction cross-section: the correlation
  seems to suggest that $\sigma^{\rm H}_{250\mu{\rm m}}$ increases
  from $1.5\times 10^{-25}$ to $2.0\times 10^{-25}$ when $T_{\rm dust}$
  decreases from 14 K to 12 K.

To estimate the effect of the uncertainty related to the adopted
emissivity index, $\beta$, and to allow direct comparison with some
previous determinations of the dust opacity (e.g.,
\citealt{2012ApJ...751...28M}, \citealt{2013ApJ...763...55R},
\citealt{2011ApJ...728..143S}), we repeated the analysis described
above using fixed emissivity indices of $\beta=1.8$ and $\beta=2.2$.
Lowering $\beta$ leads to higher dust colour temperatures, lower
optical depths $\tau_{250}$, and lower values of $\sigma^{\rm
H}_{250\mu{\rm m}}$.  The effect of increasing $\beta$ is the
opposite. The values of $\sigma^{\rm H}_{250\mu{\rm m}}$ and 
$\kappa^{\rm g}_{250\mu{\rm m}}$ resulting from a linear fit to the $H-K_{\rm s}$ vs. 
$\tau_{250 \mu{\rm m}}$ correlation for $\beta=1.8$ and $\beta=2.2$ are the 
following:
$$
\begin{array}{lclcl}
\beta=1.8 &\; : \;&   
\sigma^{\rm H}_{250\mu{\rm m}} &=& (1.5\pm0.3)\times10^{-25}\, 
{\rm cm}^2 \, {\rm H}^{-1} \; , \\
          &       &   
\kappa^{\rm g}_{250\mu{\rm m}} &=& (0.06\pm0.01)  \, {\rm cm}^2 {\rm g}^{-1} \; , \\
          &       &   & &  \\
\beta=2.2 &\; : \; &  
\sigma^{\rm H}_{250\mu{\rm m}} &=& (2.4\pm0.4)\times10^{-25}\, 
{\rm cm}^2 \, {\rm H}^{-1} \; , \\
         &         & 
\kappa^{\rm g}_{250\mu{\rm m}} &=& (0.10\pm0.02)  \, {\rm cm}^2 {\rm g}^{-1} \; .
\end{array}
$$
In both cases the extrapolated dust absorption cross-section per 
unit mass of gas at $\lambda=850\,\mu$m happens to be the same as for 
$\beta=2.0$, i.e., 
$\kappa_{850\mu{\rm m}}^{\rm g} = 0.007  \, {\rm cm}^2 {\rm g}^{-1}$.  
The values at $450\,\mu$m  change slightly depending on $\beta$:  
$\kappa_{450\mu{\rm m}}^{\rm g} = 0.022 - 0.028 \, {\rm cm}^2 {\rm g}^{-1}$ 
for $\beta=1.8-2.2$. These results should be compared with predictions from
various models listed in Table~2 of \citet{2005ApJ...632..982S}.

The suggested temperature dependence of the cross-section $\sigma^{\rm
H}_{250\mu{\rm m}}$ steepens with the adopted $\beta$.  For
$\beta=1.8$ the increase in $\sigma^{\rm H}_{250\mu{\rm m}}$ is about
24\% when $T_{\rm dust}$ decreases from 14 K to 12 K, whereas for
$\beta=2.2$ the corresponding increase is 37\%.

\section{Discussion}

The extinction ratio $A_{250 \mu{\rm m}}/A_J=0.0014\pm0.0002$, and the
implied values of the dust absorption cross-section, $\sigma_{250
  \mu{\rm m}}^{\rm H}=(1.8\pm0.3)\times10^{-25}\, {\rm cm}^2 \, {\rm
  H}^{-1}$, and the opacity per unit mass of gas, $\kappa_{250 \mu{\rm
    m}}^{\rm g} = (0.08\pm0.01) \; {\rm cm}^2 {\rm g}^{-1}$, derived
for the cold core CrA C are close to the widely adopted ``standard''
values from \citet{1983QJRAS..24..267H}. The derived absorption
  cross-section is similar or slightly smaller than the values derived
  previously for molecular clouds (e.g., \citealt{2012ApJ...751...28M}
  and references therein). For example, \cite{2009ApJ...696.1918T} determined 
  the opacity $\kappa_{160u{\rm m}}=(0.23\pm0.05)\,{\rm cm^2g^{-1}}$  at
  $160\mu{\rm m}$ (assuming $\beta=2.0$) for low extinction   regions ($A_V$ up to 4 mag) in the Taurus cloud L1521, when we get
  $\kappa_{160\mu{\rm m}}^{\rm g} = (0.19\pm0.03) \, {\rm cm}^2 {\rm
    g}^{-1}$ in CrA~C.  Adopting the spectral emissivity index
  $\beta=1.8$ used in several previous studies, we derive a dust absorption
  cross-section of $\sigma_{250 \mu{\rm m}}^{\rm
    H}=(1.5\pm0.3)\times 10^{-25}\, {\rm cm}^2 \, {\rm H}^{-1}$. This
  value lies close to the low end of the range derived
  by \cite{2012ApJ...751...28M}, using similar methods and the same
  $\beta$, towards the Vela cloud near the Galactic
  plane. We note that our $\sigma_{250 \mu{\rm m}}^{\rm H}$
has the same meaning as $\sigma_{\rm e}(1200)$, and that our
$\kappa_{250 \mu{\rm m}}^{\rm g}$ is denoted by $r\kappa_0$ in Martin
et al.  Furthermore, we have adopted the colour excess to column
density ratio from \cite{2012ApJ...751...28M}. The $\tau_{250}/N_{\rm H}$ ratio (implying $\sigma^{\rm H}_{250\mu{\rm m}} = (2.3\pm0.3)\times10^{-25}\, {\rm
    cm}^2 \, {\rm H}^{-1}$), derived by \cite{2011A&A...536A..25P} for
  the ``dense'' component (regions detected in CO) of the Taurus
  Molecular Cloud complex, is larger than the value
  obtained here assuming  $\beta=1.8$.

A rough agreement is found with some previous determinatons of the
dust opacity in dense cores. The value of $\sigma^{\rm H}_{250\mu{\rm m}}$ we obtain is similar to that determined in the
Thumbprint Nebula \citep{1998A&A...333..702L}, and midway the values
derived towards two positions in LDN 1688
\citep{2013MNRAS.428.2617R}. \citet{2011ApJ...728..143S} determined in
the protostellar core B335 the opacity ratio $\kappa_{850\mu{\rm
    m}}/\kappa_K\sim (3.2-4.8) \, 10^{-4}$, which agrees with the
ratio $\kappa_{850\mu{\rm m}}/\kappa_V \sim 4 \, 10^{-5}$ derived by
\citet{2003A&A...399L..43B} for the starless globule B68. The opacity
ratios determined by \citet{2003A&A...399.1073K} in the IC 5146
filament show more variety, $\kappa_{850\mu{\rm m}}/\kappa_V \sim
(1.3-5.0) \, 10^{-5}$, but there is a clear temperature dependence,
and CrA C should be compared with the cool parts of IC 5146, i.e. the
high end of Kramer's opacity range.

When compared with the frequently used dust model of
\citet{1994A&A...291..943O}, our average dust opacity
$\kappa_{850\mu{\rm m}}^{\rm g} = 0.007 {\rm cm}^2 {\rm g}^{-1}$ of
gas ($\sim \kappa_{850\mu{\rm m}}^{\rm d} = 0.7 {\rm cm}^2 {\rm
  g}^{-1}$ of dust) corresponds roughly to the opacity of unprocessed
dust grains taken from the MRN \citep{1977ApJ...217..425M} size
distribution, but lies clearly below the predictions for coagulated
or/and ice-coated grains of the same model (see also the comparison 
between different models in \citealt{2005ApJ...632..982S}). 

\begin{figure}[htb]
  \resizebox{\hsize}{!}
{\includegraphics{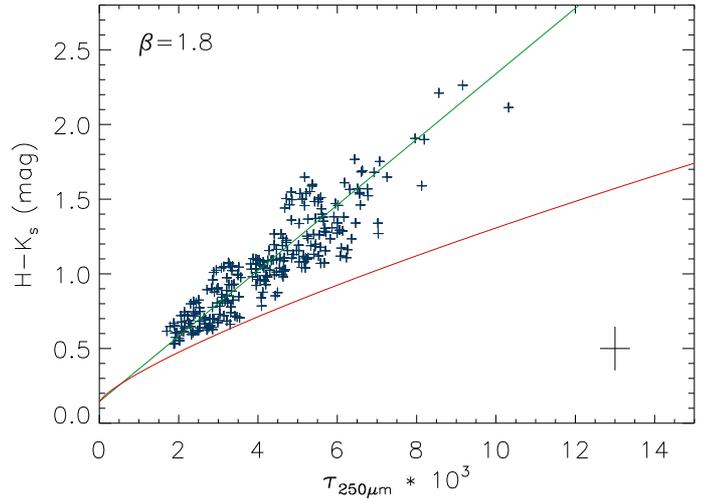}}
  \caption{$H-K$ colours of the background stars as a function 
   $\tau_{250\mu{\rm m}}$ from \textit{Herschel} assuming a dust emissivity 
   index of $\beta=1.8$. A linear fit to the data points is shown 
   with a green line. The red curve shows the relationship implied by 
   the dependence of $\sigma_{\rm FIR}^{\rm H}$ on $N_{\rm H}$ found in Orion A by 
  \cite{2013ApJ...763...55R}. The mean error of the data points is 
  indicated with a cross in the bottom right.}
\label{figure:H-Kvstau_roy} 
\end{figure}

In terms of data and methods used our study is similar to the recent
work of \cite{2013ApJ...763...55R} who determined the dust opacities
in the Orion A cloud using \textit{Herschel} and 2MASS. The study
  of \cite{2013ApJ...763...55R} benefits from a large number of data
  points, majority of them corresponding to moderate extinctions
  ($A_{\rm V} < 8$ mag). The authors found indications of an increase
in the dust opacity towards higher column densities, with 
  $(\sigma_{\rm
  FIR}^{\rm H}/10^{-25}{\rm cm^2}) = (N_{\rm H}/10^{21}{\rm cm^{-2}})^{0.28}$ (see 
their Fig. 4).  In order to
compare this relationship with the situation observed in CrA~C, we
have replotted in Fig.~\ref{figure:H-Kvstau_roy} the $H-K_{\rm s}$
versus $\tau_{250\mu{\rm m}}$ correlation derived using $\beta=1.8$ as
assumed in \cite{2013ApJ...763...55R}. In this figure, we have also
plotted the relationship corresponding to the dependence of
$\sigma_{\rm FIR}^{\rm H}$ on $N_{\rm H}$ found in Orion A, namely
$H-K_{\rm s} = (H-K_{\rm s})_0\, + \, 42\, \tau_{250\mu{\rm
    m}}^{1/1.28}$.  The change of the dust opacity towards higher
column densities seems clearly less marked in CrAC than could have
been expected on the basis of the results of
\cite{2013ApJ...763...55R} in Orion.

The $A_{\rm FIR}/A_J$ vs. $T_{\rm D}$ correlation plot suggests,
however, a change of $\sigma_{\rm FIR}^{\rm H}$ as a function of
temperature (which in turn depends on the column density), and
therefore a slight curvature can be present in the $H-K_{\rm s}$
vs. $\tau_{250\,\mu{\rm m}}$ relationship.  The slope of the $A_{\rm
  FIR}/A_J$ vs. $T_{\rm D}$ correlation is similar to that derived in
L1642 by \citet{2007A&A...466..969L} for roughly the same temperature
range as observed here, but clearly steeper than the gradient in
IC 5146 reported by \citet{2003A&A...399.1073K} (a factor of three
increase in $A_{850\mu m}/A_V$ from $T_{\rm dust}=20$ K to 12 K). The
data of Lehtinen et al. suggest that the gradient in the $A_{\rm
    FIR}/A_{\rm NIR}$ ratio becomes smoother with increasing colour
temperature. The implied increase in $\sigma_{\rm FIR}^{\rm H}$ with a
decreasing temperature possibly reflects the fact that in
starless cores the highest densities are associated with the
lowest dust temperatures. Monte Carlo simulations showed,
  however, that if the true relation is flat, noise tends to produce a
  strong negative correlation between the observed values of
  $A_{250\mu{\rm m}}/A_{\rm J}$ and T$_{\rm d}$. Assuming the 7\%
  relative uncertainty for the surface brightness measurements, the
  slope is in the simulations negative and almost three times as large
  as the observed one.  The 7\% error estimate is very conservative
  for the band-to-band errors in Herschel data. Nevertheless, we
  clearly do not have strong evidence that $A_{250\mu{\rm m}}/A_{\rm
    J}$ is a decreasing function of temperature in CrA~C.

\vspace{1mm}

Because of line-of-sight temperature variations, the fitted colour
temperature overestimates the mass-averaged dust temperature (e.g.,
\citealt{Shetty2009a}; \citealt{2012A&A...547A..11N};
\citealt{2012A&A...542A..21Y}).  This leads to an underestimation of
the optical depth, $\tau_\lambda$, and the
dust opacity $\kappa_\lambda$ towards the densest regions
(\citealt{2012A&A...539A..71J}; \citealt{Malinen2011}).

To estimate the importance of this effect in the present case, we
examined a model that closely resembles the CrA C core.  The
  model is detailed in Appendix \ref{appendix:cloudmodel}. The
  modelling result suggests that indeed, the values of $\tau_{\rm 250
    \mu m}$ derived from the observations may underestimate the
  ``true'' optical depths of the model by $\sim$30\% at the high
  end of the $H-K_{\rm s}$ versus $\tau$ correlation.  In Appendix
  \ref{appendix:sigmamodi} we examine the change in the $H-K_{\rm s}$ versus 
   $\tau_{\rm 250 \mu m}$ correlation assuming that the relationship
  between the ``observed'' and ``true'' optical depth is the same
  as in the model. It turns out that the correlation plot $H-K_{\rm s}$ versus
  the ``corrected'' $\tau_{\rm 250 \mu m}$ agrees quite well with the
  relationship implied by the dependence of $\sigma^{\rm H}$ on
  $N_{\rm H}$ determined by \cite{2013ApJ...763...55R}, namely that
  $E(H-K_{\rm s})=C \tau_{250\mu{\rm m}}^{1/1.28}$. At the high end of the
  optical depth range examined here, the values of $\sigma^{\rm
    H}_{250\mu{\rm m}}$ calculated from this relationship are
  $\sim40\%$ higher than obtained from the linear fit to the
  unmodified optical depths derived from observations. 

  To summarise, although we do not see direct evidence for changes of
  the dust opacity in our observations, it is likely that the dust
  temperature in the core decreases inward owing to the attenuation of
  the interstellar radiation field, and the optical depths,
  $\tau_{250\mu{\rm m}}$, derived from observations underestimate the
    true values up to $30\%$ near the core centre. The bias
      caused by difference between the colour temperature and the
      mass-averaged dust temperature was not corrected in the study of
      \cite{2013ApJ...763...55R}, but probably the effect there was not
      as severe as in the present study because their data probe lower
      column density regions than that discussed here.  Taking the bias
    into account, and adopting the emissivity exponent value
    $\beta=2.0$ we estimate that the dust absorption cross section per
    unit mass of gas, $\kappa^{\rm g}_{250\mu{\rm m}}$, increases from
    $\sim 0.08\, {\rm cm}^2 {\rm g}^{-1}$ at the core edges to $\sim
    0.11\, {\rm cm}^2 {\rm g}^{-1}$ close to the centre of the
    core. We note that the very nucleus of the core is not sampled in
    the present data because no stars were detected in this part.

  The dust opacities are not particularly high compared with those
found in dense clouds (e.g., \citealt{2012ApJ...751...28M}).  In the
cold cores examined by \cite{2011A&A...527A.111J}, the highest
estimates were above $\kappa^{g}_{250\mu{\rm
m}}$=0.2\,cm$^{2}$\,g$^{-1}$ but the values obtained for starless
cores in Musca cloud were below $\kappa^{g}_{250\mu{\rm
m}}\sim$0.12\,cm$^{2}$\,g$^{-1}$ and thus similar to the values in
CrA\,C.  Owing to its location in the shielded 'tail' of the CrA
molecular cloud, near the centre of an extensive HI cloud
(\citealt{1981MNRAS.196P..29L}; \citealt{1993A&A...278..569H}), the
CrA C core is likely to be at an early stage of evolution.  Previous
molecular line observations support this idea. Firstly, several common
high-density tracers, such as N$_2$H$^+$, NH$_3$ and H$^{13}$CO, are
weak in this core, suggesting that the gas has a moderate maximum
density ($n_{\rm H_2}\sim 10^5$ cm$^{-3}$). This estimate is
consistent with the cloud model constructed in Appendix
\ref{appendix:cloudmodel} where the best fit to the \textit{Herschel}
far-infrared was obtained with a central density of $n_{\rm H}
=1.5\,10^{5}\, {\rm cm}^{-3}$.  Secondly, C$^{18}$O seems to be
undepleted and peaks, together with the high-density tracers roughly
at the same place as the thermal dust emission
\citep{2003cdsf.conf..331K}. As grain growth by accretion, and the
corresponding depletion of abundant elements in the gas-phase need
time to take effect, the relatively low value of $\kappa$ is perhaps
connected with the low degree of molecular depletion derived in CrA C,
both being indicators that suggest the relative youth of this core.

\begin{acknowledgements}
  We thank Aurora Sicilia-Aguilar for providing the HAWK-I photometry
  of CrA, and Jean-Philippe Bernard for deriving the sky brightness
  offsets for the \textit{Herschel} maps using Planck satellite data.
  Helpful discussions with Kalevi Mattila are thankfully
  acknowledged. 
  This publication makes use of data products from the Two Micron All
  Sky Survey, which is a joint project of the University of
  Massachusetts and the Infrared Processing and Analysis
  Center/California Institute of Technology, funded by the National
  Aeronautics and Space Administration and the National Science
  Foundation.
  The study has been financially supported by the Academy of Finland
  through grants 127015, 132291, 140970 and 250741, and by the
  European Community FP7-ITN Marie-Curie Programme (grant agreement
  238258).
\end{acknowledgements}

\bibliographystyle{aa} 

\bibliography{CrAC_refs}

\begin{thebibliography}{52}
\expandafter\ifx\csname natexlab\endcsname\relax\def\natexlab#1{#1}\fi

\bibitem[{{Andr{\'e}} {et~al.}(2010){Andr{\'e}}, {Men'shchikov}, {Bontemps},
  {K{\"o}nyves}, {Motte}, {Schneider}, {Didelon}, {Minier}, {Saraceno},
  {Ward-Thompson}, {di Francesco}, {White}, {Molinari}, {Testi}, {Abergel},
  {Griffin}, {Henning}, {Royer}, {Mer{\'{\i}}n}, {Vavrek}, {Attard},
  {Arzoumanian}, {Wilson}, {Ade}, {Aussel}, {Baluteau}, {Benedettini},
  {Bernard}, {Blommaert}, {Cambr{\'e}sy}, {Cox}, {di Giorgio}, {Hargrave},
  {Hennemann}, {Huang}, {Kirk}, {Krause}, {Launhardt}, {Leeks}, {Le Pennec},
  {Li}, {Martin}, {Maury}, {Olofsson}, {Omont}, {Peretto}, {Pezzuto}, {Prusti},
  {Roussel}, {Russeil}, {Sauvage}, {Sibthorpe}, {Sicilia-Aguilar}, {Spinoglio},
  {Waelkens}, {Woodcraft}, \& {Zavagno}}]{2010A&A...518L.102A}
{Andr{\'e}}, P., {Men'shchikov}, A., {Bontemps}, S., {et~al.} 2010, \aap, 518,
  L102

\bibitem[{{Arab} {et~al.}(2012){Arab}, {Abergel}, {Habart}, {Bernard-Salas},
  {Ayasso}, {Dassas}, {Martin}, \& {White}}]{2012A&A...541A..19A}
{Arab}, H., {Abergel}, A., {Habart}, E., {et~al.} 2012, \aap, 541, A19

\bibitem[{{Ascenso} {et~al.}(2007){Ascenso}, {Alves}, {Beletsky}, \&
  {Lago}}]{2007A&A...466..137A}
{Ascenso}, J., {Alves}, J., {Beletsky}, Y., \& {Lago}, M.~T.~V.~T. 2007, \aap,
  466, 137

\bibitem[{{Bertin} \& {Arnouts}(1996)}]{1996A&AS..117..393B}
{Bertin}, E. \& {Arnouts}, S. 1996, \aaps, 117, 393

\bibitem[{{Bessell} \& {Brett}(1988)}]{1988PASP..100.1134B}
{Bessell}, M.~S. \& {Brett}, J.~M. 1988, \pasp, 100, 1134

\bibitem[{{Bianchi} {et~al.}(2003){Bianchi}, {Gon{\c c}alves}, {Albrecht},
  {Caselli}, {Chini}, {Galli}, \& {Walmsley}}]{2003A&A...399L..43B}
{Bianchi}, S., {Gon{\c c}alves}, J., {Albrecht}, M., {et~al.} 2003, \aap, 399,
  L43

\bibitem[{{Bohlin} {et~al.}(1978){Bohlin}, {Savage}, \&
  {Drake}}]{1978ApJ...224..132B}
{Bohlin}, R.~C., {Savage}, B.~D., \& {Drake}, J.~F. 1978, \apj, 224, 132

\bibitem[{{Cardelli} {et~al.}(1989){Cardelli}, {Clayton}, \&
  {Mathis}}]{1989ApJ...345..245C}
{Cardelli}, J.~A., {Clayton}, G.~C., \& {Mathis}, J.~S. 1989, \apj, 345, 245

\bibitem[{{Chini} {et~al.}(2003){Chini}, {K{\"a}mpgen}, {Reipurth}, {Albrecht},
  {Kreysa}, {Lemke}, {Nielbock}, {Reichertz}, {Sievers}, \&
  {Zylka}}]{2003A&A...409..235C}
{Chini}, R., {K{\"a}mpgen}, K., {Reipurth}, B., {et~al.} 2003, \aap, 409, 235

\bibitem[{{Golay}(1974)}]{1974ASSL...41.....G}
{Golay}, M., ed. 1974, Astrophysics and Space Science Library, Vol.~41,
  {Introduction to astronomical photometry}

\bibitem[{{Griffin} {et~al.}(2010){Griffin}, {Abergel}, {Abreu}, {Ade},
  {Andr{\'e}}, {Augueres}, {Babbedge}, {Bae}, {Baillie}, {Baluteau}, {Barlow},
  {Bendo}, {Benielli}, {Bock}, {Bonhomme}, {Brisbin}, {Brockley-Blatt},
  {Caldwell}, {Cara}, {Castro-Rodriguez}, {Cerulli}, {Chanial}, {Chen},
  {Clark}, {Clements}, {Clerc}, {Coker}, {Communal}, {Conversi}, {Cox},
  {Crumb}, {Cunningham}, {Daly}, {Davis}, {de Antoni}, {Delderfield}, {Devin},
  {di Giorgio}, {Didschuns}, {Dohlen}, {Donati}, {Dowell}, {Dowell}, {Duband},
  {Dumaye}, {Emery}, {Ferlet}, {Ferrand}, {Fontignie}, {Fox}, {Franceschini},
  {Frerking}, {Fulton}, {Garcia}, {Gastaud}, {Gear}, {Glenn}, {Goizel},
  {Griffin}, {Grundy}, {Guest}, {Guillemet}, {Hargrave}, {Harwit}, {Hastings},
  {Hatziminaoglou}, {Herman}, {Hinde}, {Hristov}, {Huang}, {Imhof}, {Isaak},
  {Israelsson}, {Ivison}, {Jennings}, {Kiernan}, {King}, {Lange}, {Latter},
  {Laurent}, {Laurent}, {Leeks}, {Lellouch}, {Levenson}, {Li}, {Li},
  {Lilienthal}, {Lim}, {Liu}, {Lu}, {Madden}, {Mainetti}, {Marliani}, {McKay},
  {Mercier}, {Molinari}, {Morris}, {Moseley}, {Mulder}, {Mur}, {Naylor},
  {Nguyen}, {O'Halloran}, {Oliver}, {Olofsson}, {Olofsson}, {Orfei}, {Page},
  {Pain}, {Panuzzo}, {Papageorgiou}, {Parks}, {Parr-Burman}, {Pearce},
  {Pearson}, {P{\'e}rez-Fournon}, {Pinsard}, {Pisano}, {Podosek}, {Pohlen},
  {Polehampton}, {Pouliquen}, {Rigopoulou}, {Rizzo}, {Roseboom}, {Roussel},
  {Rowan-Robinson}, {Rownd}, {Saraceno}, {Sauvage}, {Savage}, {Savini},
  {Sawyer}, {Scharmberg}, {Schmitt}, {Schneider}, {Schulz}, {Schwartz},
  {Shafer}, {Shupe}, {Sibthorpe}, {Sidher}, {Smith}, {Smith}, {Smith},
  {Spencer}, {Stobie}, {Sudiwala}, {Sukhatme}, {Surace}, {Stevens}, {Swinyard},
  {Trichas}, {Tourette}, {Triou}, {Tseng}, {Tucker}, {Turner}, {Vaccari},
  {Valtchanov}, {Vigroux}, {Virique}, {Voellmer}, {Walker}, {Ward}, {Waskett},
  {Weilert}, {Wesson}, {White}, {Whitehouse}, {Wilson}, {Winter}, {Woodcraft},
  {Wright}, {Xu}, {Zavagno}, {Zemcov}, {Zhang}, \&
  {Zonca}}]{2010A&A...518L...3G}
{Griffin}, M.~J., {Abergel}, A., {Abreu}, A., {et~al.} 2010, \aap, 518, L3

\bibitem[{{Harju} {et~al.}(1993){Harju}, {Haikala}, {Mattila}, {Mauersberger},
  {Booth}, \& {Nordh}}]{1993A&A...278..569H}
{Harju}, J., {Haikala}, L.~K., {Mattila}, K., {et~al.} 1993, \aap, 278, 569

\bibitem[{{Hildebrand}(1983)}]{1983QJRAS..24..267H}
{Hildebrand}, R.~H. 1983, \qjras, 24, 267

\bibitem[{{Indebetouw} {et~al.}(2005){Indebetouw}, {Mathis}, {Babler}, {Meade},
  {Watson}, {Whitney}, {Wolff}, {Wolfire}, {Cohen}, {Bania}, {Benjamin},
  {Clemens}, {Dickey}, {Jackson}, {Kobulnicky}, {Marston}, {Mercer},
  {Stauffer}, {Stolovy}, \& {Churchwell}}]{2005ApJ...619..931I}
{Indebetouw}, R., {Mathis}, J.~S., {Babler}, B.~L., {et~al.} 2005, \apj, 619,
  931

\bibitem[{{Juvela} {et~al.}(2011){Juvela}, {Ristorcelli}, {Pelkonen},
  {Marshall}, {Montier}, {Bernard}, {Paladini}, {Lunttila}, {Abergel},
  {Andr{\'e}}, {Dickinson}, {Dupac}, {Malinen}, {Martin}, {McGehee}, {Pagani},
  {Ysard}, \& {Zavagno}}]{2011A&A...527A.111J}
{Juvela}, M., {Ristorcelli}, I., {Pelkonen}, V.-M., {et~al.} 2011, \aap, 527,
  A111

\bibitem[{{Juvela} \& {Ysard}(2012)}]{2012A&A...539A..71J}
{Juvela}, M. \& {Ysard}, N. 2012, \aap, 539, A71

\bibitem[{{Kontinen} {et~al.}(2003){Kontinen}, {Harju}, {Caselli},
  {Heikkil{\"a}}, \& {Walmsley}}]{2003cdsf.conf..331K}
{Kontinen}, S., {Harju}, J., {Caselli}, P., {Heikkil{\"a}}, A., \& {Walmsley},
  M. 2003, in SFChem 2002: Chemistry as a Diagnostic of Star Formation, ed.
  {C.~L.~Curry \& M.~Fich}, 331

\bibitem[{{Kramer} {et~al.}(2003){Kramer}, {Richer}, {Mookerjea}, {Alves}, \&
  {Lada}}]{2003A&A...399.1073K}
{Kramer}, C., {Richer}, J., {Mookerjea}, B., {Alves}, J., \& {Lada}, C. 2003,
  \aap, 399, 1073

\bibitem[{{Lada} {et~al.}(1994){Lada}, {Lada}, {Clemens}, \&
  {Bally}}]{ladalada1994}
{Lada}, C.~J., {Lada}, E.~A., {Clemens}, D.~P., \& {Bally}, J. 1994, \apj, 429,
  694

\bibitem[{{Lehtinen} {et~al.}(2007){Lehtinen}, {Juvela}, {Mattila}, {Lemke}, \&
  {Russeil}}]{2007A&A...466..969L}
{Lehtinen}, K., {Juvela}, M., {Mattila}, K., {Lemke}, D., \& {Russeil}, D.
  2007, \aap, 466, 969

\bibitem[{{Lehtinen} {et~al.}(1998){Lehtinen}, {Lemke}, {Mattila}, \&
  {Haikala}}]{1998A&A...333..702L}
{Lehtinen}, K., {Lemke}, D., {Mattila}, K., \& {Haikala}, L.~K. 1998, \aap,
  333, 702

\bibitem[{{Li} \& {Draine}(2001)}]{2001ApJ...554..778L}
{Li}, A. \& {Draine}, B.~T. 2001, \apj, 554, 778

\bibitem[{{Llewellyn} {et~al.}(1981){Llewellyn}, {Payne}, {Sakellis}, \&
  {Taylor}}]{1981MNRAS.196P..29L}
{Llewellyn}, R., {Payne}, P., {Sakellis}, S., \& {Taylor}, K.~N.~R. 1981,
  \mnras, 196, 29P

\bibitem[{{Malinen} {et~al.}(2011){Malinen}, {Juvela}, {Collins}, {Lunttila},
  \& {Padoan}}]{Malinen2011}
{Malinen}, J., {Juvela}, M., {Collins}, D.~C., {Lunttila}, T., \& {Padoan}, P.
  2011, \aap, 530, A101+

\bibitem[{{Martin} {et~al.}(2012){Martin}, {Roy}, {Bontemps},
  {Miville-Desch{\^e}nes}, {Ade}, {Bock}, {Chapin}, {Devlin}, {Dicker},
  {Griffin}, {Gundersen}, {Halpern}, {Hargrave}, {Hughes}, {Klein}, {Marsden},
  {Mauskopf}, {Netterfield}, {Olmi}, {Patanchon}, {Rex}, {Scott}, {Semisch},
  {Truch}, {Tucker}, {Tucker}, {Viero}, \& {Wiebe}}]{2012ApJ...751...28M}
{Martin}, P.~G., {Roy}, A., {Bontemps}, S., {et~al.} 2012, \apj, 751, 28

\bibitem[{{Mathis}(1990)}]{1990ARA&A..28...37M}
{Mathis}, J.~S. 1990, \araa, 28, 37

\bibitem[{{Mathis} {et~al.}(1983){Mathis}, {Mezger}, \&
  {Panagia}}]{1983A&A...128..212M}
{Mathis}, J.~S., {Mezger}, P.~G., \& {Panagia}, N. 1983, \aap, 128, 212

\bibitem[{{Mathis} {et~al.}(1977){Mathis}, {Rumpl}, \&
  {Nordsieck}}]{1977ApJ...217..425M}
{Mathis}, J.~S., {Rumpl}, W., \& {Nordsieck}, K.~H. 1977, \apj, 217, 425

\bibitem[{{Nielbock} {et~al.}(2012){Nielbock}, {Launhardt}, {Steinacker},
  {Stutz}, {Balog}, {Beuther}, {Bouwman}, {Henning}, {Hily-Blant},
  {Kainulainen}, {Krause}, {Linz}, {Lippok}, {Ragan}, {Risacher}, \&
  {Schmiedeke}}]{2012A&A...547A..11N}
{Nielbock}, M., {Launhardt}, R., {Steinacker}, J., {et~al.} 2012, \aap, 547,
  A11

\bibitem[{{Ormel} {et~al.}(2011){Ormel}, {Min}, {Tielens}, {Dominik}, \&
  {Paszun}}]{2011A&A...532A..43O}
{Ormel}, C.~W., {Min}, M., {Tielens}, A.~G.~G.~M., {Dominik}, C., \& {Paszun},
  D. 2011, \aap, 532, A43

\bibitem[{{Ossenkopf} \& {Henning}(1994)}]{1994A&A...291..943O}
{Ossenkopf}, V. \& {Henning}, T. 1994, \aap, 291, 943

\bibitem[{{Paradis} {et~al.}(2009){Paradis}, {Bernard}, \&
  {M{\'e}ny}}]{2009A&A...506..745P}
{Paradis}, D., {Bernard}, J.-P., \& {M{\'e}ny}, C. 2009, \aap, 506, 745

\bibitem[{{Persson} {et~al.}(1998){Persson}, {Murphy}, {Krzeminski}, {Roth}, \&
  {Rieke}}]{1998AJ....116.2475P}
{Persson}, S.~E., {Murphy}, D.~C., {Krzeminski}, W., {Roth}, M., \& {Rieke},
  M.~J. 1998, \aj, 116, 2475

\bibitem[{{Peterson} {et~al.}(2011){Peterson}, {Caratti o Garatti}, {Bourke},
  {Forbrich}, {Gutermuth}, {J{\o}rgensen}, {Allen}, {Patten}, {Dunham},
  {Harvey}, {Mer{\'{\i}}n}, {Chapman}, {Cieza}, {Huard}, {Knez}, {Prager}, \&
  {Evans}}]{2011ApJS..194...43P}
{Peterson}, D.~E., {Caratti o Garatti}, A., {Bourke}, T.~L., {et~al.} 2011,
  \apjs, 194, 43

\bibitem[{{Pilbratt} {et~al.}(2010){Pilbratt}, {Riedinger}, {Passvogel},
  {Crone}, {Doyle}, {Gageur}, {Heras}, {Jewell}, {Metcalfe}, {Ott}, \&
  {Schmidt}}]{2010A&A...518L...1P}
{Pilbratt}, G.~L., {Riedinger}, J.~R., {Passvogel}, T., {et~al.} 2010, \aap,
  518, L1

\bibitem[{{Planck Collaboration} {et~al.}(2011){Planck Collaboration},
  {Abergel}, {Ade}, {Aghanim}, {Arnaud}, {Ashdown}, {Aumont}, {Baccigalupi},
  {Balbi}, {Banday}, {Barreiro}, {Bartlett}, {Battaner}, {Benabed},
  {Beno{\^i}t}, {Bernard}, {Bersanelli}, {Bhatia}, {Bock}, {Bonaldi}, {Bond},
  {Borrill}, {Bouchet}, {Boulanger}, {Bucher}, {Burigana}, {Cabella},
  {Cardoso}, {Catalano}, {Cay{\'o}n}, {Challinor}, {Chamballu}, {Chiang},
  {Chiang}, {Christensen}, {Clements}, {Colombi}, {Couchot}, {Coulais},
  {Crill}, {Cuttaia}, {Danese}, {Davies}, {Davis}, {de Bernardis}, {de
  Gasperis}, {de Rosa}, {de Zotti}, {Delabrouille}, {Delouis}, {D{\'e}sert},
  {Dickinson}, {Dobashi}, {Donzelli}, {Dor{\'e}}, {D{\"o}rl}, {Douspis},
  {Dupac}, {Efstathiou}, {En{\ss}lin}, {Eriksen}, {Finelli}, {Forni},
  {Frailis}, {Franceschi}, {Galeotta}, {Ganga}, {Giard}, {Giardino},
  {Giraud-H{\'e}raud}, {Gonz{\'a}lez-Nuevo}, {G{\'o}rski}, {Gratton},
  {Gregorio}, {Gruppuso}, {Guillet}, {Hansen}, {Harrison},
  {Henrot-Versill{\'e}}, {Herranz}, {Hildebrandt}, {Hivon}, {Hobson}, {Holmes},
  {Hovest}, {Hoyland}, {Huffenberger}, {Jaffe}, {Jones}, {Jones}, {Juvela},
  {Keih{\"a}nen}, {Keskitalo}, {Kisner}, {Kneissl}, {Knox}, {Kurki-Suonio},
  {Lagache}, {Lamarre}, {Lasenby}, {Laureijs}, {Lawrence}, {Leach}, {Leonardi},
  {Leroy}, {Linden-V{\o}rnle}, {L{\'o}pez-Caniego}, {Lubin},
  {Mac{\'{\i}}as-P{\'e}rez}, {MacTavish}, {Maffei}, {Mandolesi}, {Mann},
  {Maris}, {Marshall}, {Martin}, {Mart{\'{\i}}nez-Gonz{\'a}lez}, {Masi},
  {Matarrese}, {Matthai}, {Mazzotta}, {McGehee}, {Meinhold}, {Melchiorri},
  {Mendes}, {Mennella}, {Mitra}, {Miville-Desch{\^e}nes}, {Moneti}, {Montier},
  {Morgante}, {Mortlock}, {Munshi}, {Murphy}, {Naselsky}, {Natoli},
  {Netterfield}, {N{\o}rgaard-Nielsen}, {Noviello}, {Novikov}, {Novikov},
  {Osborne}, {Pajot}, {Paladini}, {Pasian}, {Patanchon}, {Perdereau},
  {Perotto}, {Perrotta}, {Piacentini}, {Piat}, {Plaszczynski}, {Pointecouteau},
  {Polenta}, {Ponthieu}, {Poutanen}, {Pr{\'e}zeau}, {Prunet}, {Puget}, {Reach},
  {Rebolo}, {Reinecke}, {Renault}, {Ricciardi}, {Riller}, {Ristorcelli},
  {Rocha}, {Rosset}, {Rubi{\~n}o-Mart{\'{\i}}n}, {Rusholme}, {Sandri},
  {Santos}, {Savini}, {Scott}, {Seiffert}, {Shellard}, {Smoot}, {Starck},
  {Stivoli}, {Stolyarov}, {Sudiwala}, {Sygnet}, {Tauber}, {Terenzi},
  {Toffolatti}, {Tomasi}, {Torre}, {Tristram}, {Tuovinen}, {Umana},
  {Valenziano}, {Verstraete}, {Vielva}, {Villa}, {Vittorio}, {Wade}, {Wandelt},
  {Yvon}, {Zacchei}, \& {Zonca}}]{2011A&A...536A..25P}
{Planck Collaboration}, {Abergel}, A., {Ade}, P.~A.~R., {et~al.} 2011, \aap,
  536, A25

\bibitem[{{Poglitsch} {et~al.}(2010){Poglitsch}, {Waelkens}, {Geis},
  {Feuchtgruber}, {Vandenbussche}, {Rodriguez}, {Krause}, {Renotte}, {van
  Hoof}, {Saraceno}, {Cepa}, {Kerschbaum}, {Agn{\`e}se}, {Ali}, {Altieri},
  {Andreani}, {Augueres}, {Balog}, {Barl}, {Bauer}, {Belbachir}, {Benedettini},
  {Billot}, {Boulade}, {Bischof}, {Blommaert}, {Callut}, {Cara}, {Cerulli},
  {Cesarsky}, {Contursi}, {Creten}, {De Meester}, {Doublier}, {Doumayrou},
  {Duband}, {Exter}, {Genzel}, {Gillis}, {Gr{\"o}zinger}, {Henning},
  {Herreros}, {Huygen}, {Inguscio}, {Jakob}, {Jamar}, {Jean}, {de Jong},
  {Katterloher}, {Kiss}, {Klaas}, {Lemke}, {Lutz}, {Madden}, {Marquet},
  {Martignac}, {Mazy}, {Merken}, {Montfort}, {Morbidelli}, {M{\"u}ller},
  {Nielbock}, {Okumura}, {Orfei}, {Ottensamer}, {Pezzuto}, {Popesso},
  {Putzeys}, {Regibo}, {Reveret}, {Royer}, {Sauvage}, {Schreiber}, {Stegmaier},
  {Schmitt}, {Schubert}, {Sturm}, {Thiel}, {Tofani}, {Vavrek}, {Wetzstein},
  {Wieprecht}, \& {Wiezorrek}}]{2010A&A...518L...2P}
{Poglitsch}, A., {Waelkens}, C., {Geis}, N., {et~al.} 2010, \aap, 518, L2

\bibitem[{{Rawlings} {et~al.}(2013){Rawlings}, {Juvela}, {Lehtinen}, {Mattila},
  \& {Lemke}}]{2013MNRAS.428.2617R}
{Rawlings}, M.~G., {Juvela}, M., {Lehtinen}, K., {Mattila}, K., \& {Lemke}, D.
  2013, \mnras, 428, 2617

\bibitem[{{Roussel}(2012)}]{2012arXiv1205.2576R}
{Roussel}, H. 2012, arXiv1205.2576

\bibitem[{{Roy} {et~al.}(2013){Roy}, {Martin}, {Polychroni}, {Bontemps},
  {Abergel}, {Andr{\'e}}, {Arzoumanian}, {Di Francesco}, {Hill}, {Konyves},
  {Nguyen-Luong}, {Pezzuto}, {Schneider}, {Testi}, \&
  {White}}]{2013ApJ...763...55R}
{Roy}, A., {Martin}, P.~G., {Polychroni}, D., {et~al.} 2013, \apj, 763, 55

\bibitem[{{Shetty} {et~al.}(2009){Shetty}, {Kauffmann}, {Schnee}, {Goodman}, \&
  {Ercolano}}]{Shetty2009a}
{Shetty}, R., {Kauffmann}, J., {Schnee}, S., {Goodman}, A.~A., \& {Ercolano},
  B. 2009, \apj, 696, 2234

\bibitem[{{Shirley} {et~al.}(2011){Shirley}, {Huard}, {Pontoppidan}, {Wilner},
  {Stutz}, {Bieging}, \& {Evans}}]{2011ApJ...728..143S}
{Shirley}, Y.~L., {Huard}, T.~L., {Pontoppidan}, K.~M., {et~al.} 2011, \apj,
  728, 143

\bibitem[{{Shirley} {et~al.}(2005){Shirley}, {Nordhaus}, {Grcevich}, {Evans},
  {Rawlings}, \& {Tatematsu}}]{2005ApJ...632..982S}
{Shirley}, Y.~L., {Nordhaus}, M.~K., {Grcevich}, J.~M., {et~al.} 2005, \apj,
  632, 982

\bibitem[{{Sicilia-Aguilar} {et~al.}(2011){Sicilia-Aguilar}, {Henning},
  {Kainulainen}, \& {Roccatagliata}}]{2011ApJ...736..137S}
{Sicilia-Aguilar}, A., {Henning}, T., {Kainulainen}, J., \& {Roccatagliata}, V.
  2011, \apj, 736, 137

\bibitem[{{Skrutskie} {et~al.}(2006){Skrutskie}, {Cutri}, {Stiening},
  {Weinberg}, {Schneider}, {Carpenter}, {Beichman}, {Capps}, {Chester},
  {Elias}, {Huchra}, {Liebert}, {Lonsdale}, {Monet}, {Price}, {Seitzer},
  {Jarrett}, {Kirkpatrick}, {Gizis}, {Howard}, {Evans}, {Fowler}, {Fullmer},
  {Hurt}, {Light}, {Kopan}, {Marsh}, {McCallon}, {Tam}, {Van Dyk}, \&
  {Wheelock}}]{2006AJ....131.1163S}
{Skrutskie}, M.~F., {Cutri}, R.~M., {Stiening}, R., {et~al.} 2006, \aj, 131,
  1163

\bibitem[{{Stead} \& {Hoare}(2009)}]{2009MNRAS.400..731S}
{Stead}, J.~J. \& {Hoare}, M.~G. 2009, \mnras, 400, 731

\bibitem[{{Stepnik} {et~al.}(2003){Stepnik}, {Abergel}, {Bernard}, {Boulanger},
  {Cambr{\'e}sy}, {Giard}, {Jones}, {Lagache}, {Lamarre}, {Meny}, {Pajot}, {Le
  Peintre}, {Ristorcelli}, {Serra}, \& {Torre}}]{2003A&A...398..551S}
{Stepnik}, B., {Abergel}, A., {Bernard}, J.-P., {et~al.} 2003, \aap, 398, 551

\bibitem[{{Strai{\v z}ys} \& {Lazauskait{\.e}}(2008)}]{2008BaltA..17..277S}
{Strai{\v z}ys}, V. \& {Lazauskait{\.e}}, R. 2008, Baltic Astronomy, 17, 277

\bibitem[{{Terebey} {et~al.}(2009){Terebey}, {Fich}, {Noriega-Crespo},
  {Padgett}, {Fukagawa}, {Audard}, {Brooke}, {Carey}, {Evans}, {Guedel},
  {Hines}, {Huard}, {Knapp}, {McCabe}, {Menard}, {Monin}, \&
  {Rebull}}]{2009ApJ...696.1918T}
{Terebey}, S., {Fich}, M., {Noriega-Crespo}, A., {et~al.} 2009, \apj, 696, 1918

\bibitem[{{Vuong} {et~al.}(2003){Vuong}, {Montmerle}, {Grosso}, {Feigelson},
  {Verstraete}, \& {Ozawa}}]{2003A&A...408..581V}
{Vuong}, M.~H., {Montmerle}, T., {Grosso}, N., {et~al.} 2003, \aap, 408, 581

\bibitem[{{Weingartner} \& {Draine}(2001)}]{2001ApJ...548..296W}
{Weingartner}, J.~C. \& {Draine}, B.~T. 2001, \apj, 548, 296

\bibitem[{{Ysard} {et~al.}(2012){Ysard}, {Juvela}, {Demyk}, {Guillet},
  {Abergel}, {Bernard}, {Malinen}, {M{\'e}ny}, {Montier}, {Paradis},
  {Ristorcelli}, \& {Verstraete}}]{2012A&A...542A..21Y}
{Ysard}, N., {Juvela}, M., {Demyk}, K., {et~al.} 2012, \aap, 542, A21

\end{thebibliography}

\begin{appendix}
\section{Examining dust temperature in a cloud model of CrA C}
\label{appendix:cloudmodel}
\begin{figure}
\resizebox{\hsize}{!}{\includegraphics{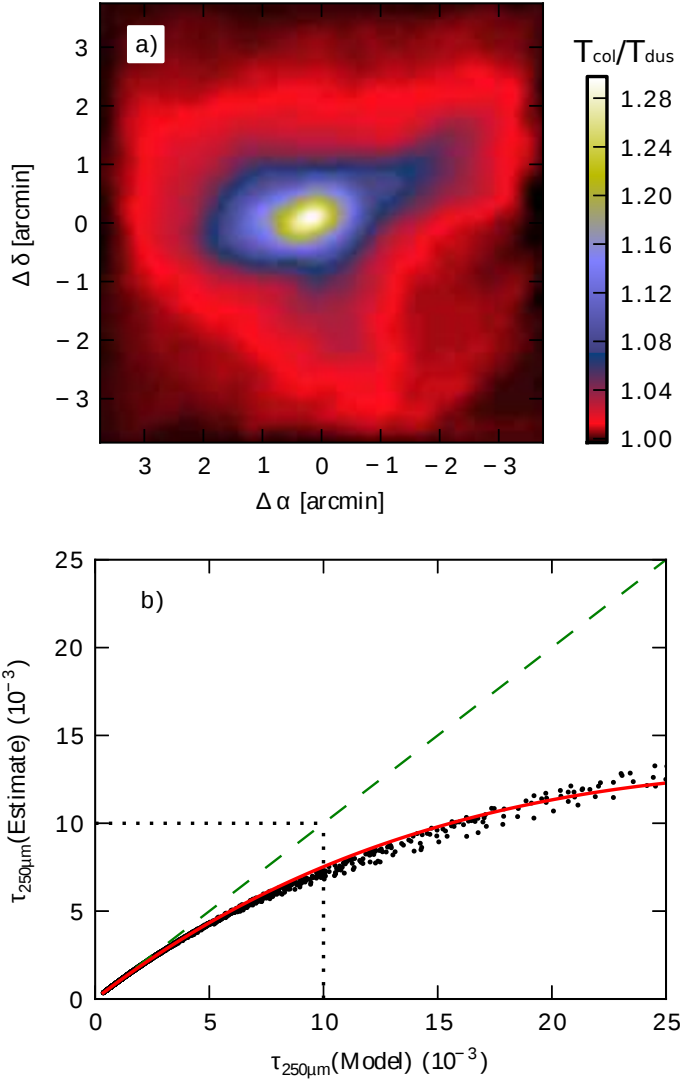}}
\caption[]{
Results from a radiative transfer model resembling the CrA C core.
{\em Frame a}: the ratio of the colour temperature (derived from
synthetic surface brightness observations) and the mass-averaged real
dust temperature. {\em Frame b}: the derived values of $\tau_{\rm
250\,\mu m}$ as the function of the true optical depth in the model.
The dotted box corresponds to the range of optical depths in 
Fig.~\ref{figure:H-Kvstau_roy}. The dashed green line represents
a 1:1 correlation. The red curve represents an 
exponential function fitted to the data points (see the text). 
}
\label{figure:model}
\end{figure}
The model cloud was discretised to 75$^3$ cells, the projected area
corresponding to $\sim 7\arcmin \times 7\arcmin$. The temperature
distribution in the three-dimensional model was solved with Monte
Carlo radiative transfer calculations, assuming the dust properties of
\cite{1994A&A...291..943O} (thin ice mantles accreted in $10^5$ years
at a density of 10$^6$). The predicted surface brightness was compared
to the data on CrA C and the column density corresponding to each of
the 75$\times$75 map pixels was adjusted iteratively until the model
exactly reproduced the 250\,$\mu$m observations. The external
radiation field was selected so that also the shape of the model
spectra and column density were similar to CrA C. To take into account
the shielding by the surrounding cloud, the external field was first
attenuated by an amount that corresponds to a dust layer with $A_{\rm
V}=1^{\rm mag}$. When the \cite{1983A&A...128..212M} field was multiplied with
a factor of 5.2, the central column density of the model was
$\sim30$\% higher than the estimated column density of CrA C and the
160\,$\mu$m intensity was $\sim$10\% above the observed.

The surface brightness maps of the model were analysed 
in the same way as the real
observations, using $\beta=1.75$ applicable for the dust model. As
expected, the column density derived from the surface brightness data
was below the true values of the model, the error being more than a
factor of two at the centre. This is illustrated in
Fig.~\ref{figure:model} where the upper frame shows the ratio of the
colour temperature and the mass-average dust temperature and the lower
frame compares the estimated and the true values of $\tau_{\rm 250\mu
  m}$.  The bias in the case of CrA C is, of course, not precisely
known but can be of similar magnitude, i.e., up to $\sim$30\% for the
highest optical depths (column densities) considered in the $H-K$
vs. $\tau_{\rm 250\mu m}$ correlation where a similar $\beta$ has been
assumed (Fig.~\ref{figure:H-Kvstau_roy}). This would decrease the
correlation slope and the reported $\sigma^{\rm H}$ and $\kappa_{\rm
  250\mu m}$ would need to be scaled up. 
Of course, similar biases may
affect most of the values reported in the literature. In what follows
we attempt to estimate the effect on $\sigma^{\rm H}$ in more detail. 

\section{Dust absorption cross section using the modified 
             optical depths}
\label{appendix:sigmamodi}

\begin{figure}
%\resizebox{\hsize}{!}
{\includegraphics[width=8cm]{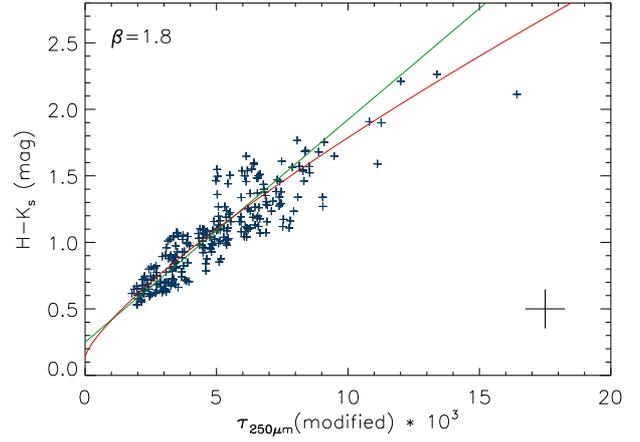}}
\caption[]{
$H-K$ colours of stars in the background of CrA~C as a function of a
modified $\tau_{250\mu{\rm m}}$ where we have attempted to correct for
the effect of line-of-sight temperature variations. It is assumed that
the relationship between the observed and ``true'' optical depth
is similar to the model shown in Fig.~\ref{figure:model}. The red curve
shows the relationship expected from the result of 
\cite{2013ApJ...763...55R} where the $\sigma^{\rm H}$ 
increases as $N_{\rm H}^{0.28}$ and the green line is a linear fit
to the data points. The mean error of the data points is 
  indicated with a cross in the bottom right.}
\label{figure:taumodel}
\end{figure}

The relationship between the ``observed'' and ``true'' optical
depths in the model can be fitted with the function $\tau_{\rm
  obs} = \tau_{\rm true}\, e^{-b \tau_{\rm true}}$, where $b=28.4$.
In Fig.~\ref{figure:taumodel}, we have plotted again the observed
$H-K$ colours in the background of CrA~C but now against a {\sl
  modified} $\tau_{250 \mu{\rm m}}$, assuming the same conversion
between the observed and ``true'' values as in the model above. In
this figure, we have also plotted the relationship 
$H-K_{\rm s} = (H-K_{\rm s})_0 + C \tau_{250\mu{\rm m}}^{1/1.28}$ implied by the 
finding of \cite{2013ApJ...763...55R} in Orion A that $\sigma^{\rm H} \propto
N_{\rm H}^{0.28}$. A reasonable agreement with the data points is
achieved by setting the constant $C=60$ instead of 42 which would make
the curve bend down too strongly. This change is equivalent with making 
the coefficient of proportionality in the relationship 
$\sigma^{\rm H} \propto N_{\rm H}^{0.28}$ smaller than in Orion by a factor of 
$(42/60)^{1.28}=0.63$. The intersection with the
y-axis is the same as in the linear fit, i.e $(H-K_{\rm
  s})_0=0.14$. A linear fit to the data, $H-K_{\rm s} = (0.25 \pm
  0.02) + (167 \pm 3)\, \tau_{250{\rm m, true}}$, is also shown in
  Fig.~\ref{figure:taumodel}. The linear fit gives a slightly larger
  $\chi^2$ than the curved relationship. Moreover, the intersection of
  the line with the $y$-axis, 0.25, is consistent with an M4 giant
  \citep{1988PASP..100.1134B} which seems less likely as a typical
  background star than a K3 giant.

Using the relationships 
$E(H-K_{\rm s}) = C \tau^{1/1.28}$, 
$\tau = N_{\rm  H}\,\sigma^{\rm H}$, and 
$N_{\rm H}=E(H-K_{\rm s})\, A(J)/E(H-K_{\rm s})\, N_{\rm H}/A(J)$ 
one obtains the following expression for the dust absorption
cross section as a funtion of the optical depth:
$$
\sigma^{\rm H}_{250\mu{\rm m}} = 
\frac{\tau^{0.28/1.28}}{C\, A(J)/E(H-K)\,N_{\rm H}/A(J)} \; . 
$$

\begin{figure}
{\includegraphics[width=8cm]{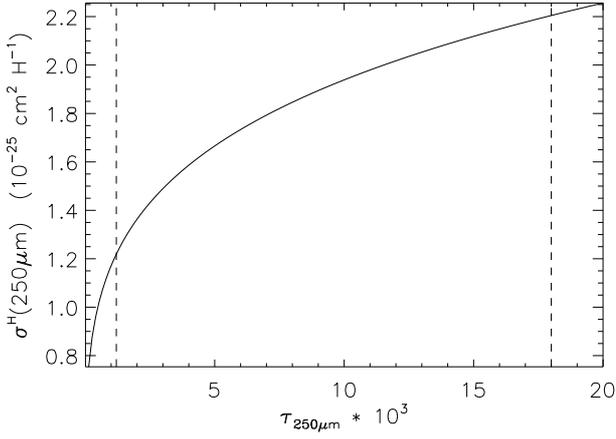}}
\caption[]{ The dust absorption cross section $\sigma^{\rm
    H}_{250\mu{\rm m}}$ as a function of $\tau_{250\mu{\rm m}}$ as
  predicted by the empirical relationship between $\sigma^{\rm H}$ and
  $N_{\rm H}$ determined by \cite{2013ApJ...763...55R} in Orion A. The
  $\tau_{250\mu{\rm m}}$ range indicated with dashed lines corresponds
  to the range of ``corrected'' optical depths in CrA~C as
  estimated on grounds of our modeling results.}
\label{figure:sigmaRoy}
\end{figure}

The dust absorption cross section resulting from this formula is shown
in Fig.~\ref{figure:sigmaRoy}, for the range of (modified)
$\tau_{250\mu{\rm m}}$ estimated for CrA~C. Here we have assumed that $C=60$. 
The model suggests that $\sigma^{\rm H}_{250\mu{\rm m}}$ increases from the value
$1.2\times10^{-25}\,{\rm cm^2 \, H^{-1}}$ to $2.2\times10^{-25}\,{\rm cm^2 \,
  H^{-1}}$ at the largest optical depths, when the linear fit
using the unmodified optical depths gave $\sigma^{\rm
  H}_{250\mu{\rm m}}= 1.5\,10^{-25}\,{\rm cm^2 \, H^{-1}}$.
  
\end{appendix}

\end{document}